\documentclass{aa}

\usepackage{graphicx}
\usepackage{booktabs}
\usepackage[utf8]{inputenc}
\usepackage{txfonts}
\usepackage[]{hyperref}
\usepackage{color, colortbl}

\begin{document}

   \title{Defunct Satellites in Nearly Polar Orbits}

   \subtitle{Long-term Evolution of Attitude Motion}

   \author{Sergey Efimov\inst{1}
          \and
          Dmitry Pritykin\inst{1}
          \and					
          Vladislav Sidorenko\inst{1}\fnmsep\inst{2}				
          }

   \institute{Moscow Institute of Physics and Technology,
              Institutskiy per., 9, 141700, Dolgoprudny, Moscow Region, Russian Federation\\
              \email{pritykin.da@mipt.ru}				
         \and				
             Keldysh Institute of Applied Mathematics, Russian Academy of Sciences,
						 Miusskaya  Sq.,  4,  125047 Moscow, Russian Federation
             }

\abstract
{Low Earth orbits (LEO) are known as a region of high space activity and, consequently, space debris highest density. Launcher upper stages and defunct satellites are the largest space debris objects, whose collisions can result in still greater pollution, rendering further space missions in LEO impossible. Thus, space debris mitigation is necessary, and LEO region is a primary target of active debris removal (ADR) projects. However, ADR planning requires at least an approximate idea of the candidate objects' attitude dynamics, which is one of the incentives for our study. This paper is mainly focused on modeling and simulating defunct satellites. The model takes into account the gravity gradient torque,  the torque due to residual magnetic moment, and the torque due to eddy currents induced by the interaction of conductive materials with the geomagnetic field. A better understanding of the intermediate phase of the exponential deceleration and existing final regimes is achieved owing to a more accurate model of the eddy currents torque than in most prior research. We also show the importance of orbital precession, which contribute to the overall attitude motion evolution.}


\maketitle

\section{Introduction}

Space debris problem is gradually becoming more notable in LEO activities. The forecast is that the situation will grow worse unless measures are taken to clear the space from the largest and the most dangerous debris objects. Different aspects of active debris removal (ADR) are brought up in (\cite{bonnal2013}). One of the generally accepted ADR scenarios is tugging debris objects to the lower orbits (\cite{aslanov2013}), whereupon they burn in the atmosphere or fall to the Earth. Most ADR techniques depend substantially on the character of the debris object's rotational dynamics, hence much effort has been spent lately to determine the rotation parameters through ground-based observations (\cite{silha2017, kucharski2014, kucharski2016}), and at the same time, much attention has been paid to studying space debris rotational dynamics analytically (\cite{praly2012, ortiz2015, linzhao2015}).

According to recent observation data published by \cite{silha2017} there are four types of objects that fall into the category of large space debris -- rocket bodies, non-functional spacecrafts, fragmentation debris, and uncorrelated objects discovered  during dedicated surveys, the former two types prevailing in numbers. Earlier we have conducted a study of the debris rocket bodies rotational dynamics (\cite{efimov2017}), and in this paper we shall focus on the defunct satellites. One of the most prominent objects in this class is, beyond doubt, Envisat satellite, whose fate is closely followed by researchers from the observation side (\cite{kucharski2014, kucharski2016}) and from the analytical side (\cite{ortiz2015}). However, as Envisat is by a wide margin the largest of the debris objects, it deserves a research in its own class, whereas in this study we shall confine ourselves to smaller, yet more representative of debris satellite population, objects.

The decay of the initial fast rotation of the large space debris objects was studied recently in \cite{praly2012, ortiz2015, linzhao2015}. We re-examine the process, using more accurate model to simulate the torque due to eddy currents as in \cite{golubkov1972, martynenko1985}, which holds true for all values of the object's angular velocity. It provides a more thorough description of the attitude motion evolution during the deceleration stage as well as the subsequent transition to final regimes. Also the orbit precession (not taken into account in \cite{praly2012, ortiz2015}) proved to have a significant influence on the properties of space debris rotational motion. One more factor that proved to be of importance in our numeric experiments is the residual magnetic moment, whose presence can produce notable qualitative changes in the variety of the final regimes.

In the second section we present our mathematical model, define all necessary reference frames, provide all the expressions for the torques contributing to the modeled dynamics, and specify the key model parameters. The third section is devoted to the numerical experiment setup, the choice of the initial conditions and parameters instrumental in interpreting the obtained results. Sections four and five present the simulation results and interpretation, the former section dealing with the ``exponential deceleration'' stage, and the latter with the classification of the final regimes and transitions to these regimes. Finally, section six relates our findings to the situation with the observation data.

\section{Mathematical model}
\subsection{Reference frames}
We shall rely on the simplest models to specify the reference frames and to describe the Earth magnetic field, because the physical factors governing the evolution of large debris rotation can be characterized in a fairly approximate manner only. We shall particularly assume the Earth to revolve with a constant velocity $\omega_E$ about an axis remaining stationary with respect to the inertial space and having a direction given by a unit vector $\mathbf{e}_E$.
We also consider only circular orbits. The center of mass (CoM) $O$ of the object in question will be taken as the origin of both reference frames.

$Oxyz$ is the body-fixed frame, whose axes coincide with the principal axes of inertia.

\begin{figure}[htb]
	\centering\includegraphics[width=3in]{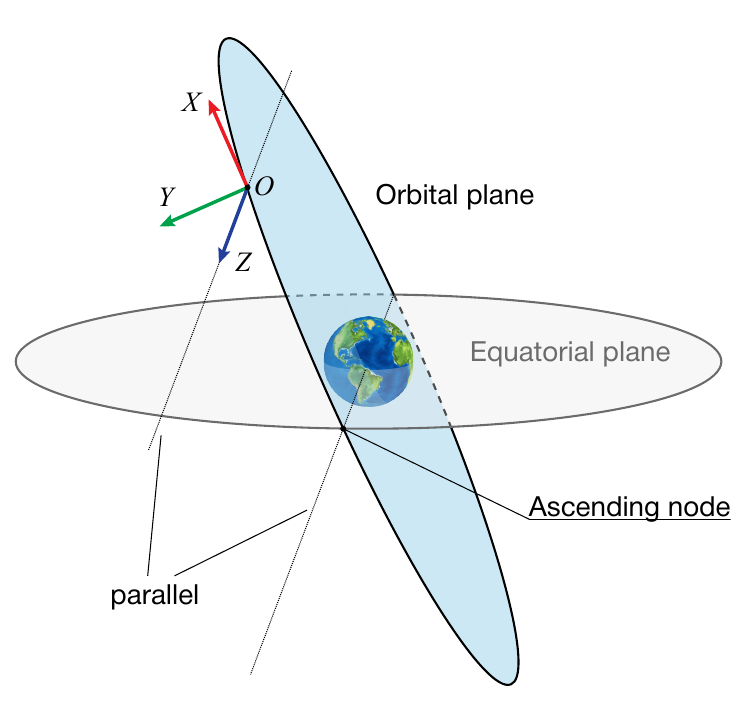}
	\caption{Semi-orbital and orbital reference frames}
	\label{figRF}
\end{figure}

$OXYZ$ is a semi-orbital reference frame: $OZ$ is parallel to the vector from the Earth's center to the ascending node, $OX$ is directed along the object's center of mass velocity as it passes the ascending node, $OY$ is perpendicular to the orbital plane (Fig.~\ref{figRF}).

\subsection{Equations of motion}

The Earth's oblateness causes the orbit's precession with angular velocity
\begin{equation*}
n_{\Omega}\approx-\frac{3 J_2 \mu_{G}^{1/2}R_E^2}{2 R_O^{7/2}}\cos{i}.
\end{equation*}
where $R_O$ and $i$ are radius and inclination of the orbit, $R_E=6378.245$~km is the Earth's mean equatorial radius, $\mu_G=3.986\cdot10^5$~km$^3$/s$^2$ is the gravity parameter of the Earth, $J_2 =1.082626\cdot10^{-3}$ is the first zonal harmonic coefficient in the expansion of the Earth's gravity field.

Argument of latitude $u$ is a linear function of time:
\begin{equation*}
\dot{u}=\omega_D,
\end{equation*}
where $\omega_D = 2\pi/T_D$, $T_D$ -- draconic period of an object's revolution around the Earth (the time between two consecutive passages through the ascending node). Employing the formula for draconic period we obtain:
\begin{equation}
\omega_D=\omega_0\left[ 1 - \frac{3}{2}{J_2}{{\left( {\frac{{{R_E}}}{R_O}} \right)}^2}(1 - 4{{\cos }^2}i) \right],
\label{eq:omegaD}
\end{equation}
where $\omega_0$ is the orbital angular velocity for the circular orbit of radius $R_O$ in the central gravity field with parameter $\mu_G$.

The rotational dynamics is described by	the Euler's equations for a rigid body motion:
\begin{equation}
\mathbf{J}_O \dot{\boldsymbol{\omega}}_B + \left[\boldsymbol{\omega}_B,\mathbf{J}_O \boldsymbol{\omega}_B\right] = \mathbf{M}_O,
\label{eq:Euler}
\end{equation}
where $\mathbf{J}_O$ is the inertia tensor, $\boldsymbol{\omega}_B$ is the object's angular velocity in the body frame, $\mathbf{M}_O$ is the vector sum of all torques acting on the object. As in prior research (\cite{praly2012, ortiz2015, linzhao2015}) when modeling the rotational dynamics with respect to object's center of mass, we shall take into account gravity gradient torque $\mathbf{M}_{G}$ and torque due to eddy currents $\mathbf{M}_{EC}$. It turns out, however, that the residual magnetic moment, given the values it achieves for the considered class of objects (up to $20 \text{ A}\cdot \text{m}^2$ according to \cite{pourtau2005}), can also contribute to the rotational dynamics evolution. Hence it will also be included into the right-hand side of \eqref{eq:Euler}. The expressions for all the torques contributing to the modeled dynamics are given below\footnote{We presume that our model describes the dynamics of the majority of defunct satellites in nearly polar orbits with altitude of 500-2000 km. It is clear, however, that finer analysis is called for in certain cases, when the motion evolution is affected by factors we left out from our study. In particular our model does not comprise the effect of secular acceleration, which has been observed in rotation of some objects  (\cite{kudak2017}).}.

The model is completed by the Poisson's kinematic equation
\begin{equation*}
\dot{q} = \frac{1}{2} q \circ \boldsymbol{\omega}_B,
\end{equation*}
where q is the object's attitude quaternion, relating the body frame to the inertial frame.

\subsection{Gravity gradient torque}
Gravity gradient torque is given by the formula (\cite{beletsky1975}):
\begin{equation}
\mathbf{M}_{G} = \frac{3\mu_G}{R_O^5} \left[\mathbf{R}_O, \mathbf{J}_O \mathbf{R}_O\right],
\end{equation}
where $\mathbf{R}_O$ is the vector from the center of the gravity field to the body's center of mass (in our case $|\mathbf{R}_O|=R_O$).

\subsection{Magnetic torque}
The torque acting upon a body with the residual magnetic moment $\boldsymbol{\mu}$ in the Earth magnetic field $\mathbf{B}$ is (\cite{shrivastava1985})
\begin{equation}
\mathbf{M}_{M} = \left[\boldsymbol{\mu}, \mathbf{B}\right].
\end{equation}

Geomagnetic field is modeled as a field of dipole placed into the center of the Earth:
\begin{equation*}
\mathbf{B} = \frac{\mu_0\mu_E}{4\pi R_O^3}\left [ \frac{3\mathbf{R}_O(\mathbf{k}_E,\mathbf{R}_O)}{R_O^2}
-\mathbf{k}_E\right],
\end{equation*}
where $\mu_0 \approx 1.257\cdot10^{-6}$~N$\cdot$A$^{-2}$ is the vacuum magnetic permeability, $\mu_E \approx 7.94\cdot10^{22}$~A$\cdot$m$^2$ is the Earth's magnetic dipole moment, $\mathbf{k}_E$ is the dipole direction. We shall assume for simplicity that the dipole is directed along the Earth's rotation axis. Test simulations showed that the use of a more precise model (inclined dipole, making an angle $\delta=11^{\circ}33'$ with the Earth's rotation axis) does not result in any qualitative difference.

\subsection{Eddy currents torque}
General formula for eddy currents torque acting on a conductive body that moves in a magnetic field can be written as follows (\cite{golubkov1972, martynenko1985}):
\begin{equation}
	\label{EC}
{{\bf{M}}_{EC}} = [\boldsymbol{F}([\boldsymbol{\omega },\bf{B}] - (\bf{v},\nabla )\bf{B} - \bf{\dot B}),\bf{B}].
\end{equation}

Here $\boldsymbol{F}$ is a magnetic tensor of a body, defined by its geometry and materials' properties. The terms $({\bf{v}},\nabla ){\bf{B}}$ and ${\bf{\dot B}}$ are related to the orbital motion of the body and the Earth's rotation. These terms are often neglected (\cite{praly2012, ortiz2015}), being small for fast rotations in comparison with $[\boldsymbol{\omega },\bf{B}]$. However, at the final stages of rotational motion evolution the angular velocity of a body is comparable to the orbital angular velocity and these terms must not be overlooked during the analysis.

\subsection{Model parameters}
\begin{figure}[htb]
	\centering\includegraphics[width=3in]{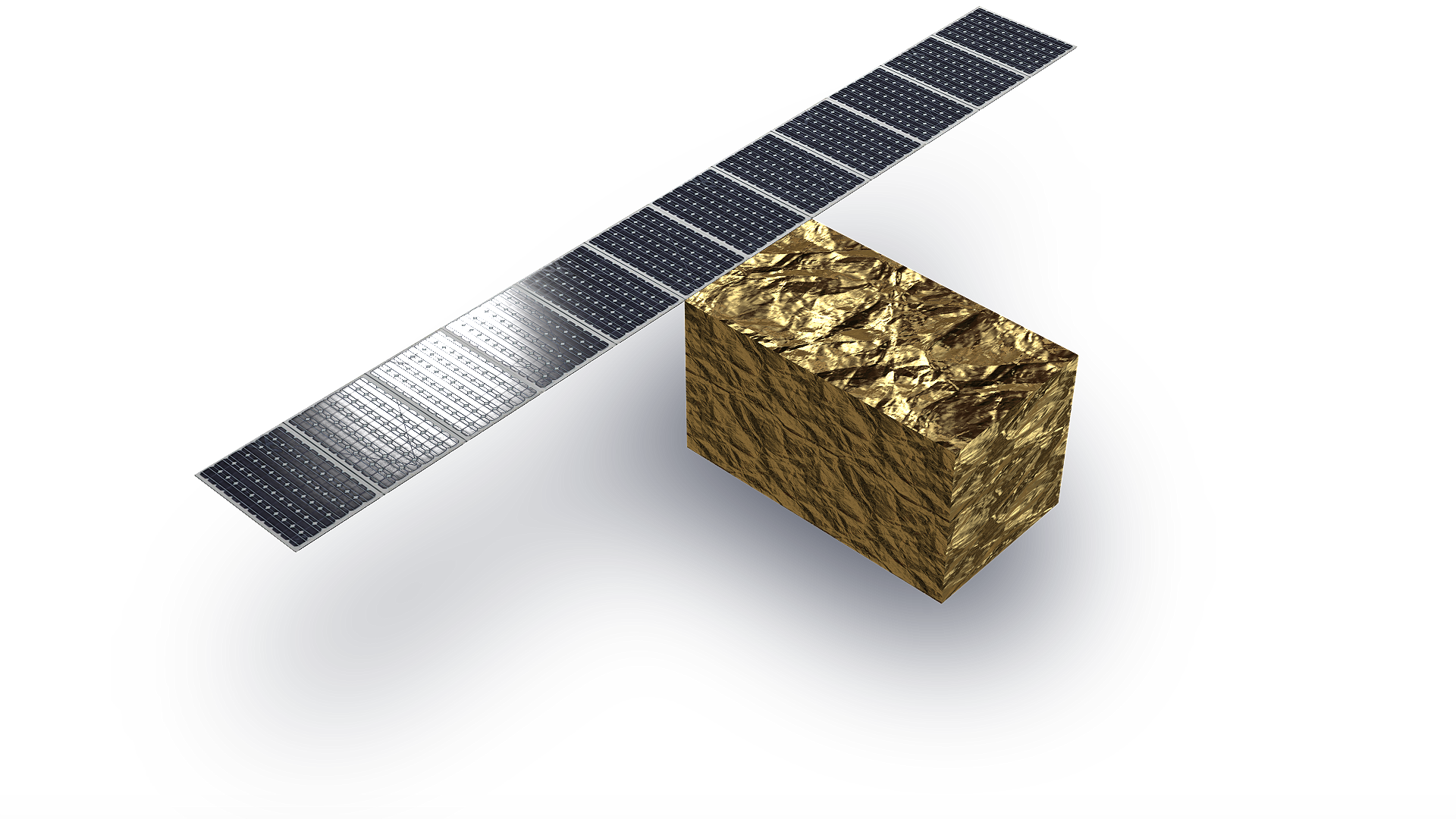}
	\caption{A schematic picture of a ``typical'' debris satellite}
	\label{figTypSat}
\end{figure}

Let us consider a ``typical'' debris satellite (Fig.~\ref{figTypSat}), whose main body linear dimensions are $2 \times 2 \times 3.5 \text{~m}$, and the dimensions of the solar panel are $16 \times 1.5 \text{~m}$. The total mass of the main body with the solar panel is $m = 1750\text{~kg}$.

The key model parameters determining the rotational dynamics are the inertia and magnetic tensors. The inertia tensor for a satellite with the specified dimensions and mass is assumed to be:
\begin{equation*}
\mathbf{J}_O=
\begin{bmatrix}
2750&0&0 \\
0&2570&0 \\
0&0&4070
\end{bmatrix} \text{kg} \cdot \text{m}^2.
\end{equation*}

We shall now estimate the magnetic tensor of such satellite. The simplest way to do so is to scale the known magnetic tensor of Envisat (\cite{ortiz2015}) in accordance with the assumed satellite dimensions. Approximating a satellite by a thin-walled conductive shell of constant thickness, one can obtain, that magnetic tensor components scale as $r^4$. This results in the following estimates of the modeled satellite magnetic tensor components
\begin{equation*}
F_1 \approx F_0 = 62.5 \cdot 10^3 \text{ S} \cdot \text{m}^4.
\end{equation*}

The tensor itself is considered to be close to spherical. Alternatively, the characteristic value of tensor components can be obtained based on the proportionality of magnetic tensor and tensor of inertia, which approximately holds in general case. Then $F \propto J$, and comparing the inertia tensor of the assumed typical satellite to that of Envisat, we obtain
\begin{equation*}
F_2 = 100 \cdot 10^3 \text{ S} \cdot \text{m}^4 \approx 2 F_0 .
\end{equation*}

These two results give an approximate understanding of the range, in which the realistic values of the modeled satellite magnetic tensor components can lie. More accurate estimates require a much finer specification of satellite design. Therefore, we  shall use both these values of magnetic tensor components -- $F_0$ and $2F_0$ -- in all subsequent simulations, covering a whole range of different possible satellite designs.

\section{Numerical Simulation Setup}

\subsection{Angles characterizing a satellite's rotational motion}
\begin{figure}[htb]
	\centering\includegraphics[width=3in]{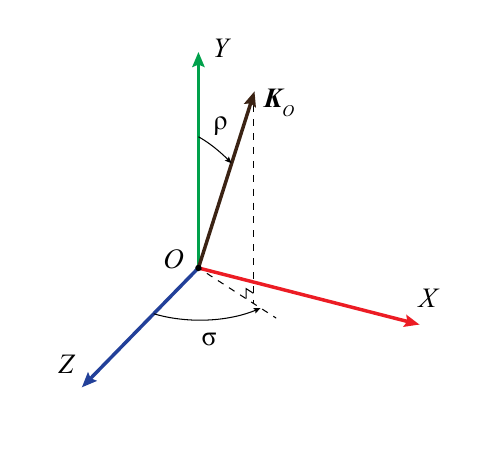}
	\caption{Angles describing the angular momentum with respect to the semi-orbital frame}
	\label{figRhoSigma}
\end{figure}

The key part in the interpretation of the attitude dynamics is played by the angular momentum vector $\boldsymbol{K}_O$. We shall describe the direction of $\boldsymbol{K}_O$ using the ``conical'' angle $\rho$ and ``hour'' angle $\sigma$ -- $\rho$ being the angle between the angular momentum and the normal to the orbit plane, whereas $\sigma$ is the angle between the direction towards the orbit’s ascending node and  $\boldsymbol{K}_O$ projection onto the osculating orbit plane (Fig.~\ref{figRhoSigma}).

We shall also use the following angles:
\begin{itemize}
\item angle $\theta$ between the axis with the minimal moment of inertia and the angular momentum;
\item angle $\delta$ between the axis with the minimal moment of inertia  and the local vertical;
\item angle $\delta_m$ between local magnetic field induction and satellite's magnetic moment. Similarly to how $\delta$ demonstrates the capture into gravitational stabilization, $\delta_m$ can reveal when a satellite becomes captured into rotation synchronous with the magnetic field rotation.
\end{itemize}

\subsection{Initial attitude motion}
It is generally agreed upon, and is corroborated by our simulations, that the attitude dynamics of a large debris object in LEO can be qualitatively divided into three major stages -- transition to ``flat'' spin due to internal dissipation, exponential deceleration due to eddy currents torque, and the stage of slow chaotic motion ending up with one of the final regimes that we shall discuss in the following sections.

Let us suppose that the modeled satellite is in the state of the fast rotation. Although modern satellites are usually equipped with the attitude control systems and preserve their orientation with respect to either inertial or orbital reference frame, fast rotations may occur as a consequence of a malfunction, that has lead to the loss of the satellite (as a well known example we can refer to the unexplained fast rotation of Envisat after failure). Since the final stages of attitude dynamics evolution revealed in our simulations are similar to motion of defunct satellites in the case of dynamically “smooth” loss of control (i.e., without sharp acceleration of rotation), our assumption is not too restrictive.

Most of the large satellites carry extended solar panels. Deformation of these panels as a satellite tumbles in the initial fast rotation stage results in internal dissipation, which causes decrease in the total kinetic energy, while the angular momentum of the system remains constant. Given the absolute value of the angular momentum, kinetic energy of a rigid body is minimal when rotational axis corresponds to the axis with the greatest moment of inertia, hence internal dissipation always transforms arbitrary rotation into ``flat'' spin (\cite{efroimsky2002}).

\begin{figure}[!h]
	\centering\includegraphics[width=3in]{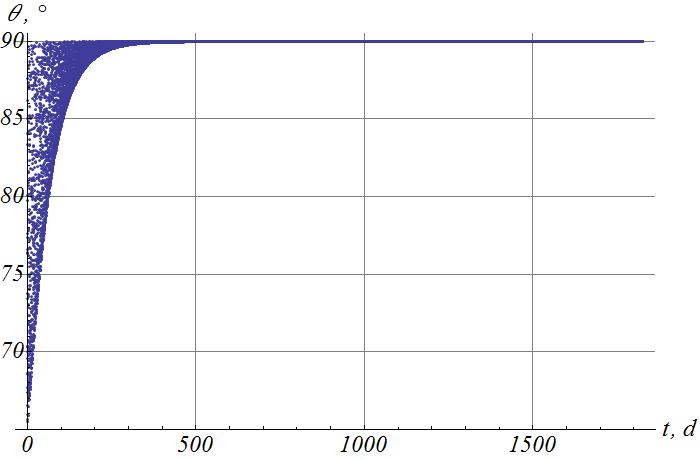}
	\caption{Transition to ``flat'' spin shown by the angle $\theta$ vs time}
	\label{figThetaSpot}
\end{figure}

An example of simulation results obtained with a model including the deformable solar panels effect is shown in Fig.~\ref{figThetaSpot}. The time of initial fast rotation transition to the ``flat'' spin mode is about one year (for $\theta(0) \approx 50^\circ$). This time is relatively short in comparison with the duration of further attitude motion evolution. For this reason, our numerical calculations will start from the point when the ``flat'' spin regime has already set in. This choice of initial conditions yields the same qualitative results for the subsequent stages of attitude dynamics evolution without complicating the mathematical model by factors that are significant only for the short transient process at the start.

The initial conditions for each series of simulations correspond to twenty different directions of the angular momentum vector uniformly distributed in space ($\rho(0)$ and $\sigma(0)$ are chosen so that the $\mathbf{K}_O$ is directed towards a dodecahedron vertexes). The absolute value of the initial angular velocity vector is $\omega(t_0)=2 \text{ deg/s}$, the modeled satellite is considered to be in the ``flat'' spin mode (rotation about the axis with the greatest moment of inertia).

\subsection{Selection of orbits and simulation parameters}
Each simulation is characterized by an orbit, a set of simulation parameters, and a set of initial conditions.

We carried out numerical experiments for three kinds of nearly polar LEO orbits -- retrograde Sun-synchronous orbit ($i=98.3^{\circ}$), polar orbit ($i=90^{\circ}$), and a prograde orbit ($i=81.7^{\circ}$) -- all with the same altitude of 770~km.

A set of simulation parameters includes the specific values of the magnetic tensor components (either  $F_0=62.5\cdot10^3 \text{ S}\cdot \text{m}^4$ or $2F_0$), the residual magnetic moment absolute value (either  $\mu_S = 10 \text{ A}\cdot \text{m}^2$ or $2 \mu_S$) and direction with respect to body frame:
\begin{align}\label{eq:mu}
\boldsymbol{\mu}_x &= \mu_S \cdot \mathbf{e}_x, \quad \boldsymbol{\mu}_y = \mu_S \cdot \mathbf{e}_y, \quad \boldsymbol{\mu}_z = \mu_S \cdot \mathbf{e}_z, \nonumber \\
\quad \boldsymbol{\mu}_d &= \frac{1}{\sqrt{3}}\mu_S\cdot\left(\mathbf{e}_x+\mathbf{e}_y+\mathbf{e}_z \right),\\
 \nonumber
\end{align}
where $\boldsymbol{e}_x$, $\boldsymbol{e}_y$, and $\boldsymbol{e}_z$ are the unit-vectors of the body-frame axes.

Our primary interest was in studying the SSO, and we conducted the simulations for all possible combinations of $\boldsymbol{\mu}$ and $\bf{F}$ (16 sets of simulations, each set containing 20 initial directions of $\boldsymbol{K}_O$). To this we added a nearly polar prograde orbit, whose inclination is symmetrical to the inclination of SSO with respect to $90^\circ$ and carried out three more sets of simulations. Finally, we conducted one simulation for the polar orbit.

\section{Simulation results: Exponential deceleration}
Our simulation results pertain to the last two major stages of the overall attitude dynamics out of the three (leaving out the short-term transition from fast rotation to ``flat'' spin). This section presents the results for the exponential deceleration stage, whereas the final regimes are discussed in the following section.

\begin{figure}[htb]
	\centering\includegraphics[width=3in]{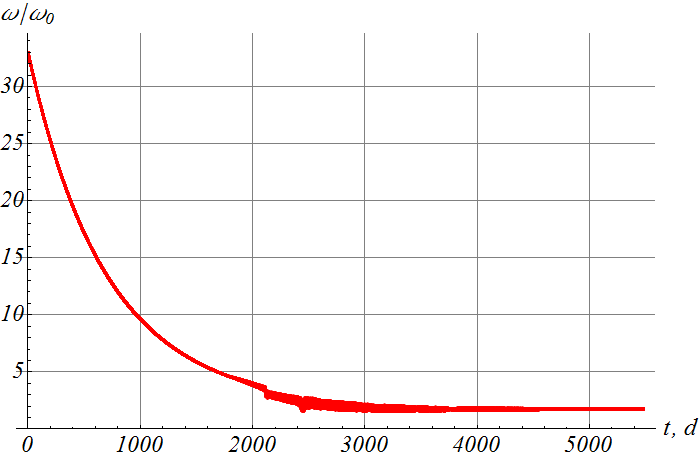}
	\caption{Angular velocity to orbital angular velocity ratio $\omega/\omega_0$ in the exponential deceleration stage}
	\label{figOmegaExpo}
\end{figure}

The stage of exponential deceleration governed by the eddy currents torque is virtually not influenced by residual magnetic moment. The value of magnetic torque averaged along the orbital motion is close to zero, which is explained by the symmetry of the geomagnetic field. Thus, characteristic times of this stage are determined primarily by magnetic tensor. For magnetic tensor with components $F_0$, the duration of this stage is
\begin{equation}
\tau \approx 1400 \text{ to } 3000 \text{ days} \approx 3.7 \text{ to } 8.2 \text{ years},
\end{equation}
and for tensor with components $2F_0$:
\begin{equation}
\tau \approx 700 \text{ to } 1500 \text{ days} \approx 1.9 \text{ to } 4.1 \text{ years}.
\end{equation}
The exact value in each case depends on the initial direction of the angular momentum vector. An example of the angular velocity to orbital angular velocity ratio $\omega/\omega_0$ dynamics is shown in Fig.~\ref{figOmegaExpo}. The angular velocity evolution, which we ascribe to the exponential deceleration stage, takes about 2100 days and its end is marked with an abrupt change in the behavior of $\omega/\omega_0$.

The regimes of motion, which set in after the exponential deceleration, substantially depend on the magnitude and direction of the residual magnetic dipole moment with respect to the body-frame. These regimes are described in the following section. The most probable of these regimes is irregular motion, which does not exhibit regular behavior in the simulation interval. Nonetheless, the absolute value of angular velocity in the final regimes does not exceed $3 \omega_0 \approx 0.18 \text{ deg/s}$.

During the exponential deceleration stage, the direction of the rotation axis changes under the combined influence of the gravity gradient and eddy currents torques. The space of $\rho$ and $\sigma$ initial values can be separated into three regions, according to the type of the subsequent angular momentum vector evolution. These regions can be loosely defined by conditions $\rho\lesssim 90^\circ$, $\rho\sim 90^\circ$, and $\rho\gtrsim 90^\circ$, and will be referred to as \emph{upper}, \emph{middle} and \emph{lower} regions respectively, because of their positions with respect to the orbital plane and the orbital normal. A detailed description of these regions' geometry is provided in \cite{efimov2017}.

There is only one type of evolution for all simulations starting from the upper region, the same is true for the middle region, whereas the initial conditions from the lower region may entail one of the two specific for this region evolution types. This yields a total of four different scenarios of evolution:
\begin{enumerate}
 \item When starting in the upper region ($\rho\lesssim 90^\circ$), the angular momentum vector at first leans towards the orbital plane, but afterwards begins to set back towards the orbital normal. This is manifested in the concave shape of $\rho$ vs time plot (Fig.~\ref{figUpperRho}). Such non-monotonous behavior is governed by the interplay of two factors. The first factor is the dissipative term $\left[\boldsymbol{\omega},\bf{B}\right]$ in \eqref{EC}, which tends to align the angular momentum vector with orbital plane. The second factor is the influence of the orbital motion on the eddy currents torque, which is described by term $\left(\bf{v},\boldsymbol{\nabla}\right)\bf{B}$ in \eqref{EC} and spins up a satellite about axis $Y$, thus pulling the angular momentum vector closer to the orbital normal. The dissipative term dominates at first, but when the orbital velocity diminishes to the value approximately 20--30 times larger than the orbital angular velocity, the orbital term starts to have a decisive effect on the angular momentum direction.
	
\begin{figure}[htb]
	\centering\includegraphics[width=3in]{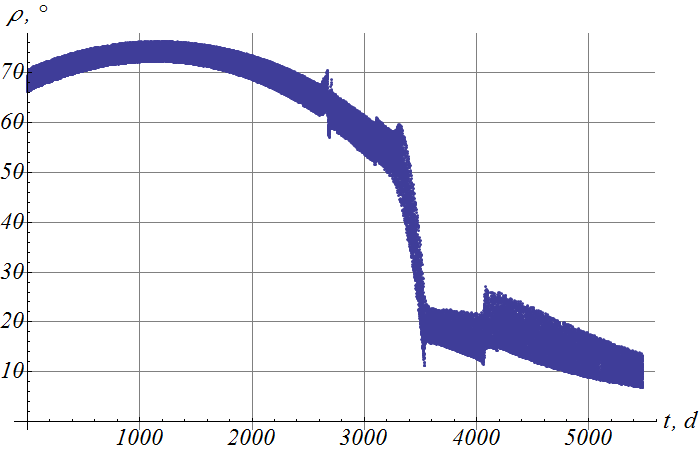}
	\caption{Evolution of angle $\rho$ in scenario 1}
	\label{figUpperRho}
	
	\centering\includegraphics[width=3in]{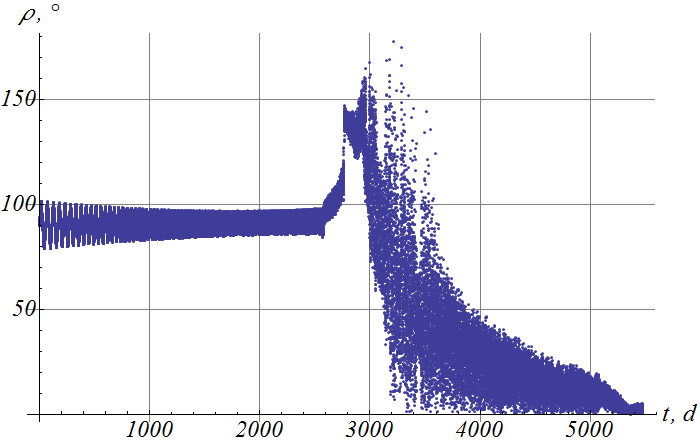}
	\caption{Evolution of angle $\rho$ in scenario 2}
	\label{figMiddleRho}
	
	\centering\includegraphics[width=3in]{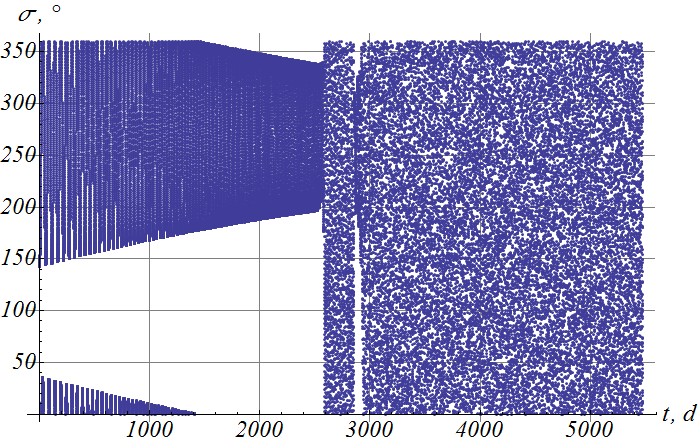}
	\caption{Evolution of angle $\sigma$ in scenario 2}
	\label{figMiddleSigma}		
\end{figure}	
	
	\item The middle region ($\rho\sim 90^\circ$) corresponds to the capture of the angular momentum vector into oscillations about $\rho=90^\circ$ (Fig.~\ref{figMiddleRho}). Fig.~\ref{figMiddleSigma} shows that the angle $\sigma$ at the same time oscillates about $\sigma=270^\circ$ for SSO (which should be the case for any other retrograde orbits) or about $\sigma=90^\circ$  for prograde orbits. For nearly polar orbits these values correspond to the approximate directions to the south celestial pole and the north celestial pole respectively. The existence of this scenario and the middle region itself is a direct consequence of the orbital plane precession.
	
	\item In the lower region ($\rho\gtrsim 90^\circ$) both dissipative and orbital terms in the eddy currents torque drive the angular momentum vector towards the orbital plane. Therefore, starting in this region the angular momentum vector monotonously approaches the orbital plane. When crossing the orbital plane, it can be captured into the oscillations similar to those in the second scenario. Overall dependence of $\rho(t)$ will thus have the shape shown in Fig.~\ref{figCaptureRho}. The capture can also be discerned in the plot of the angle $\sigma$ vs time (Fig.~\ref{figCaptureSigma}) at $t=1400~\text{d}$ as the circulation of $\sigma$ over whole interval $[0^\circ,360^\circ]$ changes to the oscillations about $\sigma=270^\circ$ with amplitude decreasing over time.

\begin{figure}[!hb]			
	\centering\includegraphics[width=3in]{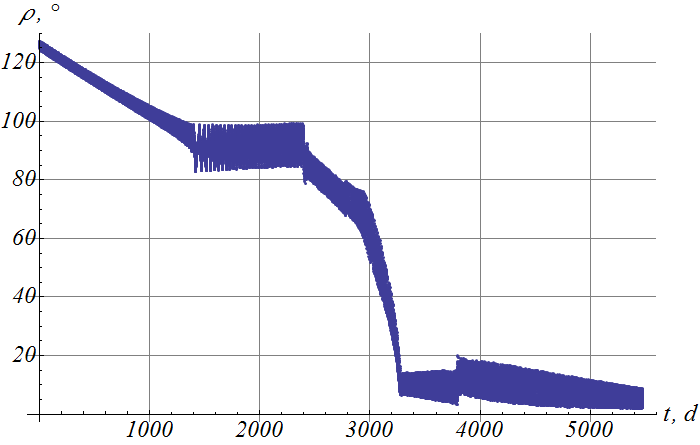}
	\caption{Evolution of angle $\rho$ in scenario 3}
	\label{figCaptureRho}
	
	\centering\includegraphics[width=3in]{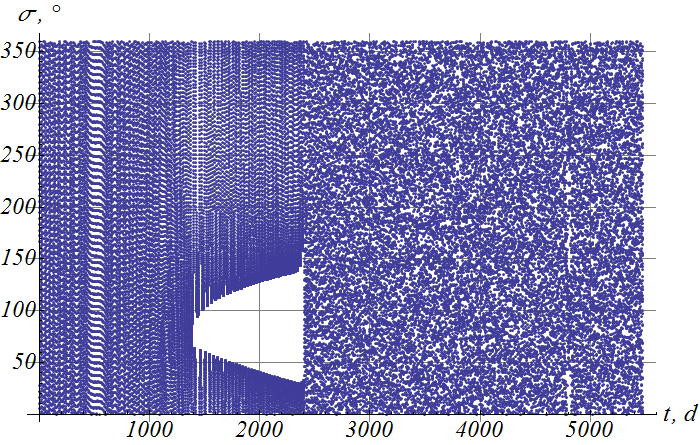}
	\caption{Evolution of angle $\sigma$ in scenario 3}
	\label{figCaptureSigma}	
	
	\centering\includegraphics[width=3in]{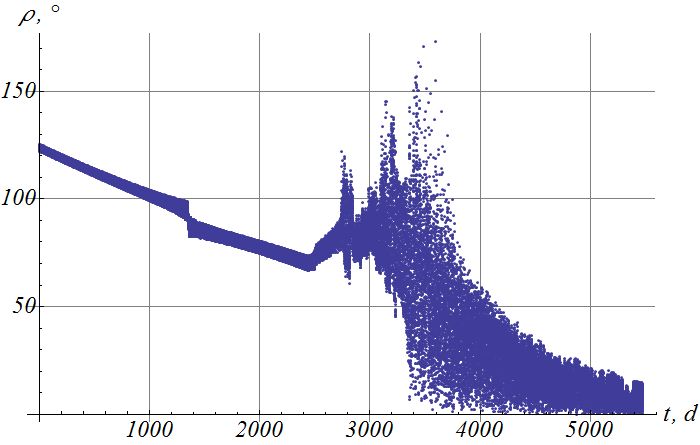}
	\caption{Evolution of angle $\rho$ in scenario 4}
	\label{figTransitRho}		
\end{figure}		
	
	\item	Alternatively, after approaching the orbital plane, the angular momentum vector can jump past it to the upper region (see the abrupt change of $\rho$ in Fig~\ref{figTransitRho} when it crosses the $90^\circ$ at $t \approx 1300~\text{ d}$), as the orbital term in the eddy currents torque drives the angular momentum vector towards the orbital normal, equivalently to the latter part of scenario 1 (Fig.~\ref{figTransitRho}). The choice between scenarios 3 and 4 has a quasi-probabilistic nature, as very close sets of initial conditions can result in different behaviors of the angular momentum vector near the orbital plane.
\end{enumerate}

\section{Simulation results: final regimes}
Six qualitatively different types of motion, which settles in the system after the stage of exponential deceleration, are observed in our numerical experiments. We shall now briefly describe all six regimes, show the characteristic graphs, and then -- at the end of this section -- we shall present the statistical distribution of the final regimes in our simulations.

\subsection{Gravitational stabilization}
	
\begin{figure}[htb]
	\centering\includegraphics[width=3in]{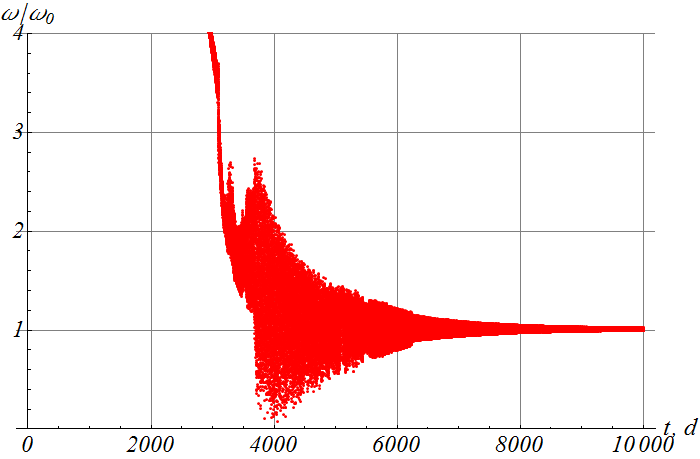}
	\caption{Angular velocity to orbital angular velocity ratio $\omega/\omega_0$ in case of gravitational stabilization final regime}
	\label{figOmegaGrav}
\end{figure}

In this regime, the axis with the smallest moment of inertia is aligned along the local vertical. Thus, the object rotates with the angular velocity $\omega = \omega_0 \approx 0.06 \text{ deg/s}$ about orbital normal. The graph of the $\omega/\omega_0$ is shown in Fig.~\ref{figOmegaGrav}.
	
This regime, however, does not take place in any of the numerical experiments with non-zero residual magnetic moment, as for chosen values of residual magnetic moment, the magnetic torque dominates over the gravity gradient torque. The non-existence of the gravitational stabilization regime in the presence of the magnetic torque is an essential difference issuing from comprising the magnetic torque into the model.

On the contrary, in experiments with zero magnetic moment 5 out of 20 simulations ended up with the gravitational stabilization.

\subsection{Magnetic rotation}

\begin{figure}[htb]
	\centering\includegraphics[width=3in]{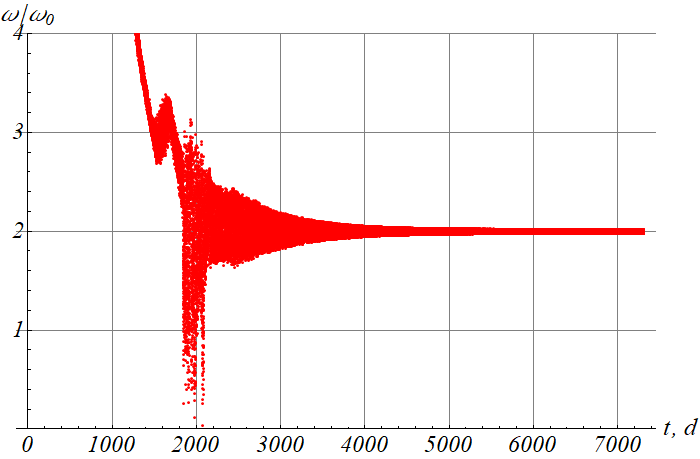}
	\caption{Angular velocity to orbital angular velocity ratio $\omega/\omega_0$ in case of magnetic rotation final regime}
	\label{figOmegaMagn}
\end{figure}

In this regime, satellite rotates synchronously to geomagnetic field vector in the semi-orbital reference frame. The angular velocity, therefore, is $\omega = 2 \omega_0 \approx 0.12 \rm{ deg/s}$ and directed along the orbital plane normal. An example evolution of $\omega/\omega_0$ is shown in Fig.~\ref{figOmegaMagn}. This regime takes place more often in cases with magnetic moment directed along axes $x$ and $y$ or for $\boldsymbol{\mu}||\boldsymbol{\mu}_d$, appearing in about 30\% of these simulations. When the magnetic moment is directed along the axis with the greatest moment of inertia, this regime is replaced by eddy currents rotation (see Section \ref{sec:1.8}).

\subsection{Magnetic tumbling}

\begin{figure}[!htb]
    \centering\includegraphics[width=3.5in]{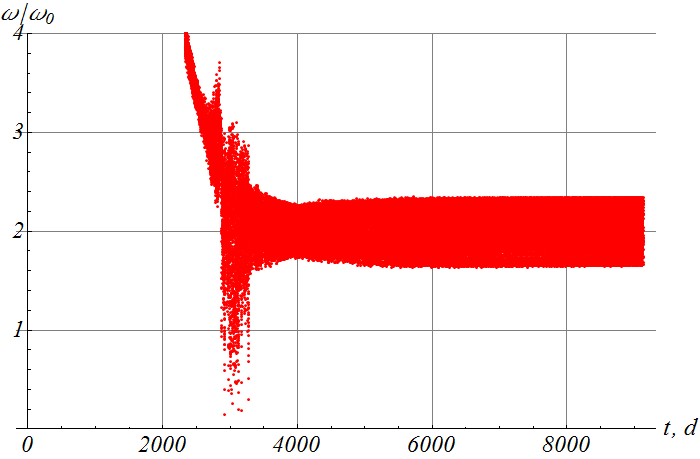}
    \caption{Angular velocity to orbital angular velocity ratio $\omega/\omega_0$ in magnetic tumbling.}
	\label{figOmegaTumbl}

	\centering\includegraphics[width=3.5in]{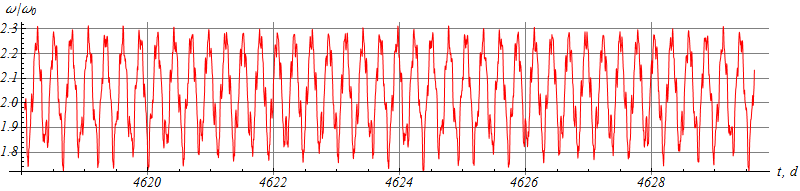}
	\caption{Quasi-periodic variations of  $\omega/\omega_0$ in magnetic tumbling (a close up view of the plot in Fig.~\ref{figOmegaTumbl}) }
	\label{figOmegaTumbl_macro}

	\centering\includegraphics[width=3.5in]{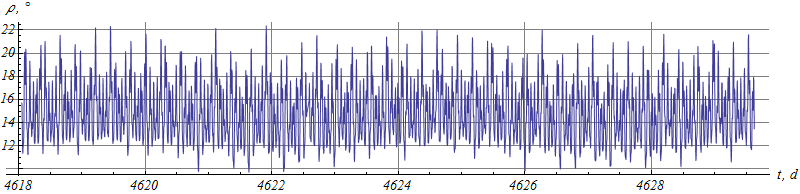}
	\caption{Variation of conical angle $\rho$ between the angular momentum and orbital normal in magnetic tumbling}
	\label{figRhoTumbl}

	\centering\includegraphics[width=3.5in]{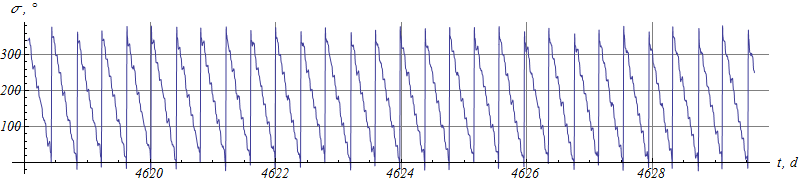}
	\caption{Variation of the hour angle $\sigma$  in magnetic tumbling }
	\label{figSigmaTumbl}

	\centering\includegraphics[width=3.5in]{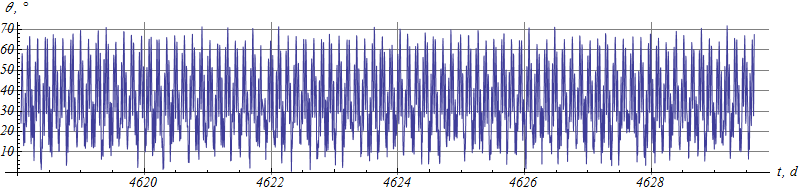}
	\caption{Variation of the angle $\theta$ between the axis with the minimal moment of inertia and the angular momentum in magnetic tumbling}
	\label{figThetaTumbl}
\end{figure}

This is a rare particular case of the magnetic rotation regime. In this case the average angular velocity is also close to $\omega = 2 \omega_0$, however, it oscillates about this value with amplitude $~0.5\omega_0 \approx 0.03 \text{ deg/s}$ instead of being virtually constant (Fig.~\ref{figOmegaTumbl}-\ref{figOmegaTumbl_macro}). The angles that describe the attitude motion also change quasi-periodically about their mean values (Fig.~\ref{figRhoTumbl}-\ref{figSigmaTumbl}). Particularly, the angle $\theta$  -- between the angular momentum vector and  the minimal moment of inertis axis -- varies in the range of $\left[10^{\circ},70^{\circ}\right]$ (Fig.~\ref{figThetaTumbl}). Thus, the motion in the magnetic tumbling regime is significantly different from ``flat rotation'' in case of magnetic rotation, and can be considered as an intermediate case between magnetic rotation and chaos, which is to be described in Section \ref{sec:chaos}.

\subsection{Eddy currents rotation, resonance 9:5} \label{sec:1.8}

For slowly rotating objects a combination of the gravity gradient torque and dissipation caused by the eddy currents tends to align the angular momentum vector along the orbital plane normal. Since the change in the angular momentum of a body caused by the eddy currents torque over the time span of one orbital period is very small, the torque Eq.~\eqref{EC} can be averaged along the spin and orbital motion. Hence, one can obtain the equilibrium value of angular velocity $\omega$ to orbital angular velocity $\omega_0$ ratio:
\begin{equation}
	\label{1.8}
\omega /{\omega _0} = 9/5=1.8.
\end{equation}

This condition represents the final regime alternative to gravitational stabilization. We have already established that in the simulations without the residual magnetic moment 5 out of 20 experiments ended up with the gravity stabilization regime. The other 15 experiments out of these 20 came to the eddy current rotation mode. The existence of similar regime has also been indicated in \cite{martynenko1985}.

It may also be noted, that this final regime cannot occur for objects in geosynchronous equatorial orbit, as they do not move with respect to the Earth's magnetic field and therefore in Eq.~\eqref{EC} $({\bf{v}},\nabla){\bf{B}}={\bf{\dot B}}=0$.

In the simulations with the non-zero magnetic moment, the regime is present only when the magnetic moment is directed along the axis with the greatest moment of inertia, appearing in almost 60\% of such numeric experiments. The reason behind this is that the axis with the greatest moment of inertia is the initial axis of rotation, due to transient process in the beginning of evolution. Consequently, magnetic torque is perpendicular to the axis of rotation; therefore it does not influence angular velocity value and does not prevent the satellite from being captured into the eddy currents rotation, unless the axis of rotation is significantly altered during the stage of slow chaotic motion.

\begin{figure}[htb]
	\centering\includegraphics[width=3in]{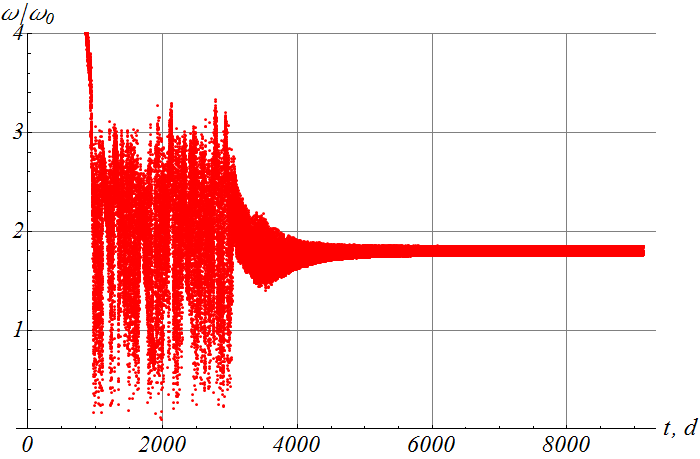}
	\caption{Angular velocity to orbital angular velocity ratio $\omega/\omega_0$ in case of eddy currents rotation final regime}
	\label{figOmega18}
\end{figure}

\subsection{Chaotic dynamics}\label{sec:chaos}

In this scenario the stage of slow chaotic motion following the exponential deceleration does not end in foreseeable future (25-40~years from the beginning of the simulation), and thus never comes to one of four previously described regimes. Even in cases, where motion converges to one of the regular final regimes, the influence of residual magnetic moment can significantly prolong the transient stage of the irregular rotational dynamics. E.g., the characteristic duration of the transition from the exponential deceleration stage to a final regime without magnetic moment is about 2 years, whereas with the magnetic moment directed along axis $x$ it can lasts for up to 7 years.

\begin{figure}[htb]
	\centering\includegraphics[width=3in]{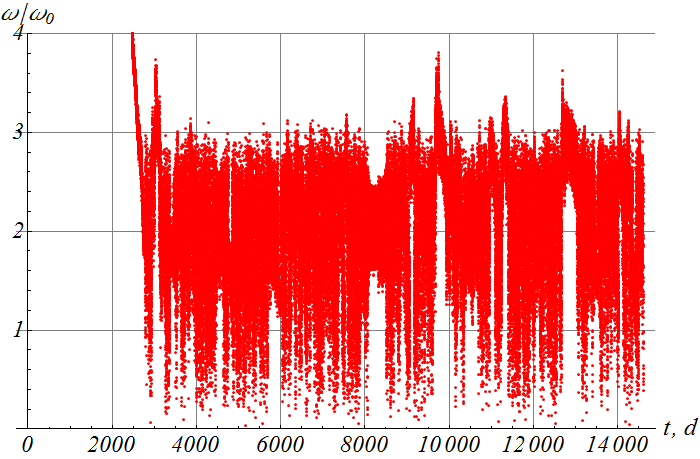}
	\caption{Angular velocity to orbital angular velocity ratio $\omega/\omega_0$ in case of chaotic final regime}
	\label{figOmegaChaos}
\end{figure}

The angular velocity during the chaotic motion fluctuates in the range  $0 \text{ to } 3\omega_0 \approx 0.18 \text{ deg/s}$ as shown in Fig.~\ref{figOmegaChaos}.

For the specified magnetic moment values, the chaotic motion is very common, as about 65\% of all simulations end up with this regime. There are two cases, however, in which this probability is significantly lower than its average. In experiments with $(\mu;F)=(\mu_S;2F_0)$ the ratio of dissipative eddy currents torque to magnetic torque reaches its maximum value, which helps to stabilize the motion. Consequently, chaotic motion appears as a final regime in less than 40\% of these simulations. Also in the case of residual magnetic moment directed along the axis with the greatest moment of inertia, as described earlier, magnetic torque does not directly oppose the capture in eddy currents rotation and the probability of chaotic final motion is decreased to approximately 40\%. The opposite worst case is represented by simulation, in which $(\boldsymbol{\mu};F)=(2\boldsymbol{\mu}_y;F_0)$. Here the dissipation is minimal, the magnetic moment is maximal and directed along the axis with the smallest moment of inertia. As a result, 19 out of 20 experiments end up in chaotic motion.

It is interesting to note that the possibility of chaotic regimes of motion of a magnetized satellite has long been known to specialists (e.g., \cite{beletsky1999, liqun2003}). And yet the observational evidence of such regimes is still lacking.

\subsection{Resonance 19:10?}

\begin{figure}[htb]
	\centering\includegraphics[width=3in]{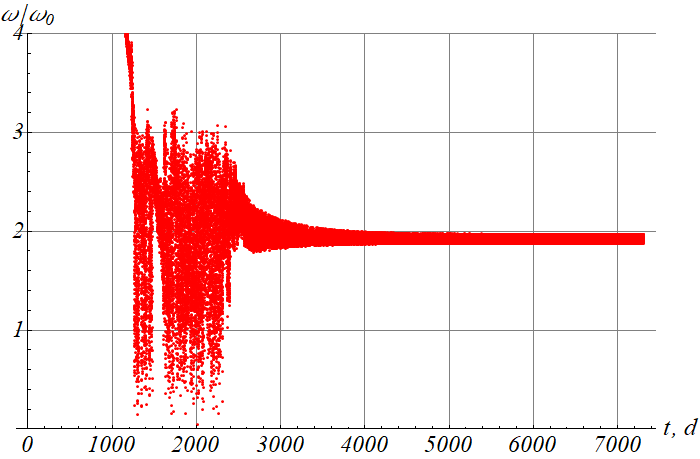}
	\caption{Angular velocity to orbital angular velocity ratio $\omega/\omega_0$ in case of resonance 19:10}
	\label{figOmegaPolar}
\end{figure}

This regime was observed in the simulations for the polar orbit, although there are reasons to believe that it was also present in simulations for other orbits, although was not classified as such. The polar orbit simulation was conducted for a pair $(2\boldsymbol{\mu}_z; 2 F_0)$. The new final regime appears in 3 cases out of 20. The angular velocity in this regime stabilizes around some constant value, but this value is neither $1.8\omega_0$ nor $2 \omega_0$ (Fig.~\ref{figOmegaPolar}). Thus, this regime cannot be identified as either magnetic rotation or eddy current rotation.

\begin{figure}[htb]
	\centering\includegraphics[width=3in]{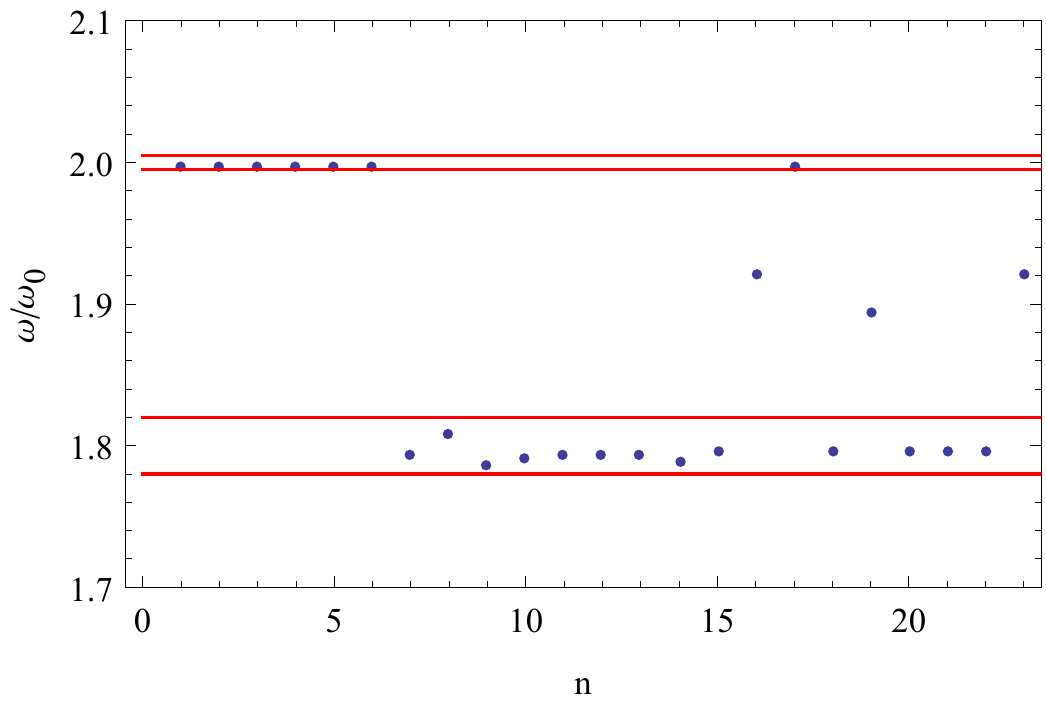}
	\caption{Mean value of angular velocity in final regimes for different simulations. Red lines envelop points, that can be identified as close to the eddy current rotation regime $1.8\omega_0$ and the magnetic rotation regime $2.0\omega_0$ }
	\label{figFinReg}
\end{figure}

This can be clearly seen in Fig.~\ref{figFinReg}, which depicts the mean value of angular velocity in the final stage for different simulations. Three dots that cluster outside of the narrow bands around $1.8 \omega_0$ and $2 \omega_0$ pertain to the new final regime. The mean value of the angular velocity in the final stage for this regime is very close to $1.9 \omega_0$. Therefore, it can be assumed to be a hybrid regime, governed by the joint effect of eddy currents and torque due to residual magnetization, although exact mechanism behind it is not clear.

\begin{figure}[htb]
	\centering\includegraphics[width=3in]{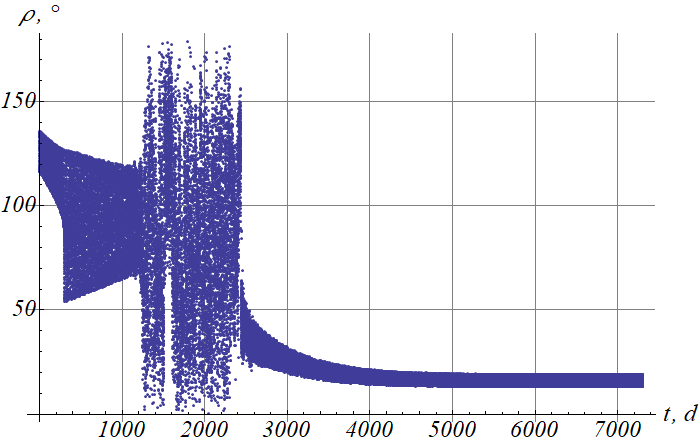}
	\caption{Variation of angle $\rho$ in case of resonance 19:10}
	\label{figRhoPolar}
	
	\centering\includegraphics[width=3in]{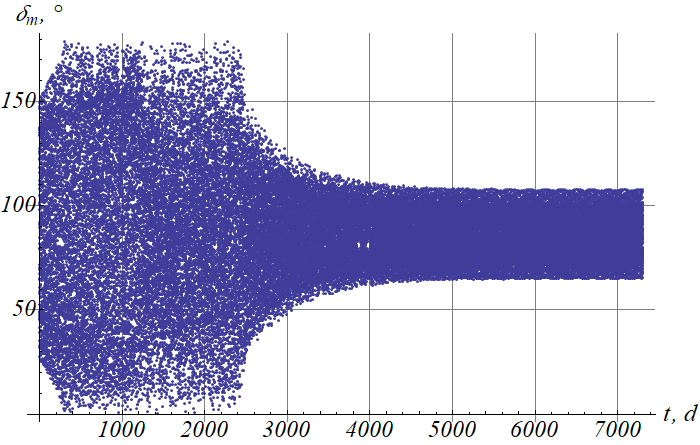}
	\caption{Variation of angle $\delta_m$ in case of resonance 19:10}
	\label{figDeltaPolar}	
\end{figure}

Unlike the cases of the magnetic rotation and eddy currents regimes the angle $\rho$ in the new regime does not converge to zero (Fig.~\ref{figRhoPolar}), and the rotation axis precess about orbital normal. The angle $\delta_m$ oscillates about $90^\circ$ with large amplitude (Fig.~\ref{figDeltaPolar}) instead of very small amplitude as in the case of the eddy currents regime or convergence to $0^\circ$ for magnetic rotation regime.

\subsection{Final regimes statistics}

Tables 1–3 provide the information on the numbers of cases that results in each type of different final regimes in the SSO simulations. Rows are sorted in descending order of residual magnetic moment to magnetic tensor components ratio. Columns are labeled by the indices of magnetic moment vectors \eqref{eq:mu} and sorted by the angle that the residual magnetic dipole moment makes with the axis with the greatest moment of inertia (secondary criterion being the angle it makes with the axis with the smallest moment of inertia). Hereby in accordance with previous description of the chaotic final regime, the number of chaotic cases in Table~1 decreases from top to bottom and from left to right.

The total number of simulations in Tables 1–3 is 320 (16 combinations of $\boldsymbol{\mu}$ and $\bf{F}$, 20 initial conditions for each combination). We can see from the data that 203 cases out of 320 end up with the chaotic final regime, 46 out of 320 arrive at the eddy currents rotation (resonance 9:5), and the rest 71 come to the magnetic rotation (6 out of them to magnetic tumbling). Let us recall that out of 20 simulations with the absence of the residual magnetic moment 5 cases rested in the gravitational stabilization mode, whereas the other 15 came to the eddy currents rotation. It emphasizes the fact that the large (although realistic) value of the residual magnetic moment drastically changes the evolution of a debris satellite rotational motion.

Tables 6–7 compare the simulations made for the prograde orbit with the corresponding simulations for SSO (we recall that the simulations for the prograde orbits are confined to only three pairs of $(\boldsymbol{\mu}; F)$). The statistics on the prograde orbits is not large enough to make definite conclusions, but it does not look to differ much from the analogous SSO numbers.

Finally, Table 7 provides the comparison data of the polar orbit simulation against the corresponding SSO simulation. The polar orbit data is remarkable for the examples of an unprecedented final regime, which we called resonance 19:10 and the exact nature of which is yet to be explained.

\begin{table}[!h]
\begin{center}
\caption{SSO simulations: chaotic final motion for different sets of simulation parameters}
\begin{tabular}{ c | c c c c |c}
\toprule
  & $(2\mu_S;F_0)$ & $(\mu_S;F_0)$ & $(2\mu_S;2 F_0)$ & $(\mu_S;F_0)$ & $\Sigma$\\
\midrule	
  $y$      & 19  & 16 & 14 & 10  & 59  \\
  $x$      & 17  & 17 & 15 & 8   & 57  \\
	$d$       & 18  & 16 & 17 & 6   & 57  \\
  $z$      & 8   & 11 & 5  & 6   & 30  \\
\midrule	
  $\Sigma$ & 62  & 60 & 51 & 30  & 203 \\	
\bottomrule	
\end{tabular}

\vspace{20pt}

\caption{SSO simulations: eddy currents rotation for different sets of simulation parameters}
\begin{tabular}{ c | c c c c |c}
\toprule
  & $(2\mu_S;F_0)$ & $(\mu_S;F_0)$ & $(2\mu_S;2 F_0)$ & $(\mu_S;F_0)$ & $\Sigma$\\
\midrule	
  $y$      & 0  & 0 & 0  & 0  & 0  \\
  $x$      & 0  & 0 & 0  & 0  & 0  \\
	$d$       & 0  & 0 & 0  & 0  & 0  \\
  $z$      & 11 & 8 & 14 & 13  & 46 \\
\midrule	
  $\Sigma$ & 11 & 8 & 14 & 13 & 46 \\	
\bottomrule	
\end{tabular}

\vspace{20pt}

\caption{SSO simulations: magnetic rotation (including magnetic tumbling) for different sets of simulation parameters. The number of magnetic tumbling cases, when present, is provided in brackets}
\begin{tabular}{ c | c c c c |c}
\toprule
  & $(2\mu_S;F_0)$ & $(\mu_S;F_0)$ & $(2\mu_S;2 F_0)$ & $(\mu_S;F_0)$ & $\Sigma$\\
\midrule	
  $y$      & 1    & 4     & 6  & 10    & 21    \\
  $x$      & 3    & 3     & 5  & 12(2) & 23(2) \\
	$d$       & 2(2) & 4(1)  & 3  & 14(1) & 23(4) \\
  $z$      & 1    & 1     & 1  & 1     & 4     \\
\midrule	
  $\Sigma$ & 7(2) & 12(1) & 15 & 37(3) & 71(6) \\	
\bottomrule	
\end{tabular}
\end{center}
\end{table}

\begin{table}[!h]
\begin{center}
\caption{Prograde orbit and SSO simulations comparison: chaotic final motion}
\begin{tabular}{ l c c c }
\toprule
    & $(2\boldsymbol{\mu}_y;F_0)$ & $(\boldsymbol{\mu}_x;2F_0)$ & $(\boldsymbol{\mu}_z;F_0)$\\
\midrule	
  Sun-synchronous orbit  & 19 & 8  & 11 \\
  Prograde orbit  & 20 & 14 & 12 \\
\bottomrule	
\end{tabular}

\vspace{20pt}

\caption{Prograde orbit and SSO simulations comparison: eddy currents rotation}
\begin{tabular}{ l c c c }
\toprule
    & $(2\boldsymbol{\mu}_y;F_0)$ & $(\boldsymbol{\mu}_x;2F_0)$ & $(\boldsymbol{\mu}_z;F_0)$\\
\midrule	
  Sun-synchronous orbit  & 0 & 0  & 8 \\
  Prograde orbit  & 0 & 0 & 8 \\
\bottomrule	
\end{tabular}

\vspace{20pt}

\caption{Prograde orbit and SSO simulations comparison: magnetic rotation (including magnetic tumbling). The number of magnetic tumbling cases, when present, is provided in brackets}
\begin{tabular}{ l c c c }
\toprule
    & $(2\boldsymbol{\mu}_y;F_0)$ & $(\boldsymbol{\mu}_x;2F_0)$ & $(\boldsymbol{\mu}_z;F_0)$\\
\midrule	
  Sun-synchronous orbit  & 1 & 12(2)  & 1 \\
  Prograde orbit  & 0 & 6 & 0 \\
\bottomrule	
\end{tabular}

\vspace{20pt}

\caption{Polar orbit and SSO simulations comparison}
\begin{tabular}{ l c c c c}
\toprule
    & Chaos & \multicolumn{1}{p{1.5cm}}{\centering Magnetic \\ rotation} & \multicolumn{1}{p{1.5cm}}{\centering Eddy currents \\ rotation} & \multicolumn{1}{p{1.5cm}}{\centering Resonance \\ 19:10}\\
\midrule	
   Polar orbit  & 7 & 1  & 9 & 3\\
   SSO   & 5 & 1 & 14 & -\\
\bottomrule	
\end{tabular}
\end{center}
\end{table}

\section{Perspectives of observational verification of predicted effects}

The current state of the art in space debris attitude motion determination is described by \cite{silha2017}. Different tools are used for observations of real objects in LEO: optical telescopes, laser rangefinders, radars, etc.  But in most cases everything is limited to optical observations. Laser and radar observations are only available for unique objects like Envisat (\cite{kucharski2014, sommer2017}).

The results of systematic optical observations of a large number of objects in LEO were summarized recently by \cite{dearborn2012}, \cite{silha2017}. Unfortunately, these observations are primarily focused on the rocket bodies. \cite{silha2017} reported that in LEO they observed 100 rocket bodies and only 15 defunct satellites. \cite{dearborn2012} did not observe satellites at all.

Nevertheless, the need to intensify the observations of objects in LEO is recognized by all specialists. This lets us hope that in the near future the data will be made available to verify the results of our simulations. In particular, very promising results are provided by intensive observations of near-Earth objects that have been going on since 2014 using the multi-channel monitoring telescope located in Nizhny Arkhyz, Russia (\cite{beskin2017})

\section{Conclusion}

Using numerical simulation we conducted a study of the attitude dynamics of a ``typical'' defunct satellite in nearly polar orbit. Our aim was to describe the ``nominal'' long-term evolution of satellite’s attitude motion, i.e. the evolution in the absence of extraordinary events like, for example, fuel leakage, collisions with fragments of the space microcosm, and partial destruction. It turns out, that this evolution can be subdivided into three stages: transient process, exponential deceleration and the stage of slow chaotic motion ending up with one of the final regimes, whose properties essentially depend on the interplay between gravity torque, magnetic torque and eddy currents torque applied to the satellite. The final regimes of defunct satellites exhibit greater diversity than the final regimes in the other class of large space debris -- rocket bodies. In addition to gravitational stabilization, which is the usual end of game for rocket bodies, the final regimes of defunct satellite attitude motion include different types of slow rotations and even a chaotic tumbling. The next challenge is to discover the predicted effects in the motion of the real objects. It may be of particular interest to check if the angular momentum vector of fast rotating space debris in nearly polar orbits can indeed oscillate about the direction to the north or south celestial pole.

\begin{acknowledgements}
Research reported in this paper was supported by RFBR (grant 17-01-00902).
\end{acknowledgements}

\end{document}